# INFLUENCE OF RIPENING AND CREAMING ON THE AGGREGATION RATE OF DODECANE-IN-WATER NANOEMULSIONS. IS THE CREAMING RATE AN APPROPRIATE MEASURE OF EMULSION STABILITY?

Yorlis Mendoza, Kareem Rahn-Chique, Antonio M. Puertas, Manuel S. Romero-Cano, German Urbina-Villalba*

*SUMMARY*

*The behavior of four oil-in-water (O/W) ioinic nanoemulsions composed of dodecane, and mixtures of dodecane with squalene and tetra-chloro-ethylene is studied. These nanoemulsions were stabilized with sodium dodecyl sulfate (SDS). The behavior of the turbidity and the average radius of the emulsions were followed as a function of time. The results illustrate the shortcomings of characterizing the stability of emulsions by their creaming rate.*

PALABRAS CLAVE: Aggregation / Creaming / Oswald / Emulsion / Nano

*RESUMEN*

*Se estudia el comportamiento de cuatro nano-emulsiones iónicas de aceite en agua (O/W) compuestas por dodecano puro y mezclas de dodecano con escualeno y tetracloroetileno. Las nanoemulsiones fueron estabilizadas con dodecil-sulfato de sodio (DSS). El comportamiento de su radio promedio y turbidez fue estudiado en función del tiempo. Los resultados ilustran las limitaciones de caracterizar la estabilidad de emulsiones a través de su tasa de formación de crema.*

PALABRAS CLAVE: Nano / Agregación / Sedimentación / Oswald / Emulsión / Stability

**Yorlis Mendoza.** Ingeniero Químico (2012) de la Universidad Nacional Experimental Francisco de Miranda. Punto Fijo, Edo. Falcón, Venezuela. Email: yorliska_05@hotmail.com

**Kareem Rahn-Chique**. Licenciado en Química de la Universidad Central de Venezuela. Profesional Asociado a la Investigación (PAI-D1) del Instituto Venezolano de Investigaciones Científicas (IVIC). Centro de Estudios Interdisciplinarios de la Física. Caracas, Venezuela. Email: krahn@ivic.gob.ve.

**Antonio M. Puertas**. Doctor en Física de la Universidad de Granada, España. Profesor Titular del Grupo de Sistemas Complejos de la Universidad de Almería. Dpto. de Física Aplicada. E-mail: apuertas@ual.es

**Manuel S. Romero-Cano** Doctor en Ciencias Físicas de la Universidad de Granada, España. Profesor Titular de la Universidad de Universidad de Almería. Dpto. de Física Aplicada. E-mail: msromero@ual.es

**German Urbina-Villalba**. Doctor en Ciencias mención Química de la Universidad Central de Venezuela. Investigador Titular del IVIC. Jefe del Laboratorio de Fisicoquímica de Coloides. Centro de Estudios Interdisciplinarios de la Física. Email: guv@ivic.gob.ve

I. INTRODUCTION

According to the Laplace equation (Evans and Wennerström, 1994), the internal pressure of a drop of oil suspended in water, is directly proportional to its interfacial tension $\gamma$, and inversely proportional to its radius ($R_i$). The interfacial tension originates a difference between the chemical potential of the molecules of oil inside the drops and the ones of an unbounded bulk oil phase. This additional free energy is equal to: $(4\pi R_i^2)\gamma/N_{m,i}$ where, $N_{m,i}$ is the number of molecules of drop "i": $(4\pi R_i^3)\rho_o N_A/3MW$ (where $\rho_o$ is the density of the oil, MW its molecular weight, and $N_A$ is Avogadro's number (6.02 x $10^{23}$ molec/mol)). Hence, the referred difference in the chemical potentials is equal to:

$$\Delta\mu \approx \frac{3\gamma V_M}{R_i} \tag{1}$$

Where $V_M$ is the molar volume of the oil. Equation (1) indicates that the molecules of oil which belong to drops of different sizes have distinct chemical potentials. This promotes the diffusive transfer of oil through the water phase, a phenomenon known as Ostwald ripening. The theory of Lifthitz, Slesov, and Wagner (LSW) (Lifshitz and Slesov, 1961; Wagner, 1961), predicts that the ripening rate can be quantified in terms of the linear increase of the cubic critical radius of the emulsion as a function of time:

$$V_{OR} = dR_c^3/dt = 4\alpha D_m C(\infty)/9 \tag{2}$$

Where $R_c$, $D_m$, $C(\infty)$ and $\alpha$ stand for the critical radius of the dispersion, the diffusion constant of the oil molecules, their bulk solubility in the presence of a planar Oil/Water (O/W) interface, and the capillary length defined as:

$$\alpha = 2\gamma V_M / \tilde{R} T \qquad (3)$$

Here $\tilde{R}$ is the universal gas constant, and $T$ the absolute temperature. Finsy (2004) demonstrated that the critical radius of the dispersion is equal to its number average radius ($R_a$):

$$R_c = R_a = \frac{1}{N_T} \sum_k R_k \qquad (4)$$

Where $N_T$ is the total number of drops. According to LSW, drops with radii smaller than the critical radius decrease in size, and vice-versa. Since the critical radius changes as a function of time, the theory predicts that all drops are constantly dissolving or growing, favoring the development of a self-similar drop size distribution (DSD) at very long times. The predictions of the theory regarding Eq. (2) and the form of the DSD correspond to the so-called *stationary regime* in which a characteristic, left-skewed, drop-size distribution with a cut-off radius of $1.5R_c$ should be attained.

As the drops of any emulsion, the drops of a nanoemulsion are also subject to a constant buoyancy force resulting from the gravity field of the earth and Arquimides´ law:

$$F_b = 4\pi R_i^3 (\rho_o - \rho_w) g / 3 \qquad (5)$$

This force causes the formation of cream at the top of the container, and generates a gradient of concentration in the number of particles per unit volume n(t) along its vertical axis. The velocity of creaming can be easily estimated from the deterministic term of the equation of motion of Emulsion Stability Simulations (ESS) (Urbina-Villalba, 2000), if the Brownian movement of the particles is neglected:

$$\Delta L = \frac{D_i F}{k_B T} \Delta t \Rightarrow V_g = \frac{\Delta L}{\Delta t} \approx \frac{[k_B T / 6 \pi \eta R_i] F_b}{k_B T} \quad (6)$$

Where $\Delta L$ is the height of the container: $\Delta L = r_i(t + \Delta t) - r_i(t)$, $r_i(t)$ the position of particle "i" at time t, $\eta$ the viscosity of the water phase, and $k_B$ the Boltzmann constant. Using Eq. (5) for the buoyancy force, Eq. (6) yields:

$$V_g = \frac{[k_B T / 6 \pi \eta R_i][4 \pi R_i^3 \Delta \rho g]}{3 k_B T} = \frac{2 R_i^2 \Delta \rho g}{9 \eta} \quad (7)$$

Assuming that the height of the container is equal to $\Delta L$ = 10 cm (typical size of the sample vessel in a spectrophotometer), a dodecane drop located at the bottom of the container will require either 17 seconds ($R_i$ = 100 μm) or 205 days ($R_i$ = 100 nm) to reach the top, depending on its size.

Notice that the potential energy of the buoyancy phenomenon does not affect the chemical potential of the oil molecules which is constant, and independent of $R_i$:

$$\frac{\Delta G_b}{N_{m,i}} = \left(\frac{4\pi R_i^3}{3}\right)(\rho_o - \rho_w)g(h_0 - h) \Big/ (4\pi R_i^3)\rho_o N_A/3MW$$

$$= (1 - \rho_w/\rho_o)g(h_0 - h)MW/N_A \qquad (8)$$

Where: $h_0$ and $h$ are the height of the liquid column and the position of the drop along its vertical axis, respectively.

According to Eq. (8) the ripening rate is not influenced by gravity. However, since the Ostwald ripening process changes the radii of the particles by molecular exchange, the creaming velocity changes due to Ostwald ripening. A similar phenomenon occurs if a set of individual drops flocculate irreversibly to form an aggregate. The total mass of the aggregate is substantially higher than the one of their drops, and therefore, the influence of the buoyancy force on the displacement of the cluster is larger. This can be appreciated if the radius of drop $i$ is substituted by the hydrodynamic radius of an aggregate of size $k$ ($R_k = \sqrt[3]{k}\,R_0$) in Eq. (5).

Moreover, since the relative movement of the drops either by convection (orthokinetic flocculation) or by diffusion (perikinetic flocculation) depends on the particle size, the flocculation rate also does. If the drops coalesce after flocculation, the critical radius of the emulsion increases and larger drops will dissolve. On the other hand, if the ripening occurs in the absence of coalescence, the DSD of the emulsion still changes favoring the occurrence of a variety of aggregates with distinct flocculation rates.

The examples given in the last two paragraphs illustrate the complex situation that might occur if a pair of destabilization process occurs simultaneously. Fortunately, the effect of the gravity field can be disregarded if the difference between the density of the phases is small (see $\Delta\rho$ in Eq. 7). Similarly the ripening process could be diminished if the chosen oil is sparingly soluble in the water phase (see $C(\infty)$ in Eq. (2)). When these conditions are met, the process of coagulation proceeds as described by Smoluchowski (1917). According to this author, the number concentration of aggregates of *k* primary particles existing at time t: $n_k$ *(t)*, results from a balance between the aggregates produced by the collisions between clusters of smaller sizes *i* and *j* (such that *i+j = k*), and the aggregates of size *k* lost by the collisions with clusters of any other size:

$$\frac{dn_k(t)}{dt} = \frac{1}{2}\sum_{i=1,j=k-i}^{i=k-1} k_{ij} n_i(t) n_j(t) - n_k(t)\sum_{i=1}^{\infty} k_{ik} n_i(t) \tag{9}$$

The kernel of Eq. (9) is the set of coagulation rates between aggregates *i* and *j*: *{$k_{ij}$}*. The evaluation of these rates is very difficult. Fortunately, the rate of doublet formation ($k_{11}$) is accessible from turbidity measurements. By definition the turbidity of a dispersion ($\tau_{exp}$) is equal to:

$$\tau_{exp} = (1/L_c)\ln(I_0/I) \tag{10}$$

where: $I_0$ is the intensity of the incident light, and *I* is the intensity of light emerging from a cell of path length $L_c$ (generally of the order of $10^{-2}$ m). Theoretically, the turbidity can be expressed as:

$$\tau_{theo} = \sum_{k=1}^{\infty} n_k \sigma_k \tag{11}$$

Where $\sigma_k$ is the optical cross section of an aggregate of size k. Thus, according to Eqs. (10) and (11), the turbidity of a dispersion measures the amount of light which is lost by the scattering of the aggregates existing in a liquid.

If the time of measurement is short enough and the dispersion is sufficiently dilute, the initial slope of the turbidity as a function of time can be directly related to the rate of doublet formation $k_{11}$ [Lips et al., 1971; Lips and Willis, 1973]:

$$\left(\frac{d\tau}{dt}\right)_0 = \left[\frac{\ln 10}{L_c}\right]\left(\frac{dAbs}{dt}\right)_0 = \left(\frac{1}{2}\sigma_2 - \sigma_1\right)k_{11} n_0^2 \qquad (12)$$

Here, Abs = log($I_0$/I) is the absorbance of the dispersion, $n_0$ = n(t = 0), and $\sigma_1$ and $\sigma_2$ are the optical cross sections of a spherical drop and a doublet. According to the Rayleigh, Gans, and Debye (RGD theory [Kerker, 1969]), the cross sections of singles and doublets can be computed using Eq. (12) whenever:

$$C_{RGD} = (4\pi a/\lambda)(m-a) << 1 \qquad (13)$$

Where $\lambda$ is the wavelength of light in the medium ($\lambda = \lambda_0/n_w$, where $n_w$ is the index of refraction of water, and $\lambda_0$ the wavelength of light in vacuum), and *m* is the relative refractive index between the particle and the surrounding medium ($n_o$/$n_w$). Whenever Eq. (13) holds:

$$\sigma_k = \frac{4}{9}\pi R_k^2 \alpha_k^4 (m-1)^2 \int_0^\pi P_k(\vartheta)(1+\cos^2\vartheta)\sin(\vartheta)d\vartheta \qquad (14)$$

Where: $\vartheta$ is the angle of observation, $\alpha_k = 2\pi R_k/\lambda$, and $P_k(\vartheta)$ is the form factor of an aggregate of size k, deduced by Puertas et al. (Puertas et al., 1997; Puertas et al., 1998; Maroto y de las Nieves, 1997). In the case of a sphere (singlet):

$$P_1(\vartheta) = \left(3\frac{\sin u - u \cos u}{u^3}\right)^2 \qquad (15)$$

with $u = 2R \sin(\vartheta/2)$. In the case of a doublet:

$$P_2(\vartheta) = \left(2 + \frac{\sin 2u}{u}\right)\left(3\frac{\sin u - u \cos u}{u^3}\right)^2 \qquad (16)$$

In this work, the stability of four different nanoemulsions is studied. This emulsions are composed of : a) dodecane (Nano A), b) dodecane ($C_{12}$) + squalene (SQ) (Nano B), c) $C_{12}$ + tetracloroethylene (TCE) (Nano C), and finally, d) $C_{12}$ + SQ + TCE (Nano D). Four types of stability measurements are used:

1. The variation of $R^3$ vs. t:

$$V_{\exp} = dR_a^3/dt \qquad (17)$$

2. The profile of the transmittance of the emulsion as function of the height of the container at for different times.

3. The creaming velocity (Eq. 7).

4. The rate of doublet formation $k_{11}$ (Eq. 12).

It is known [Kamogawa et al., 1999] that the addition of an insoluble component to the oil (like squalene) decreases its aqueous solubility. It is also recognized that mixtures of dodecane with a denser substance (like TCE) can be used to produce a neutrally buoyant combination of oil. Hence, it is expected that Nano A will experience all destabilization phenomena occurring in a typical emulsion, while in the emulsions B and C the occurrence of ripening and buoyancy, respectively, will be restricted. Since Nano D contains both TCE and SQ, *hypothetically* only flocculation and coalescence are expected.

II. EXPERIMENTAL PROCEDURE

*II. 1. Materials*

Dodecano (Aldrich, 99%, 0.75 g/cc) was eluted twice through an aluminum column in order to improve its purity. Sodium chloride NaCl (Sigma, 99.5%), iso-pentanol (Scharlau Chemie, 99%, 0.81 g/cc), tetrachloroethylene (J.T. Baker, 100%, 1.614 g/cc) Sodium dodecyl sulfate (Sigma, 99%) and squalene (Aldrich, 99%, 0.809 g/cc) were used as received. The water of the experiments was distilled and deionized (1.1 $\mu$ S/cm$^{-1}$ at 25ºC) using a Simplicity purificator from Millipore (USA).

*II. 2. Nanoemulsion Synthesis and Characterization*

Nanoemulsions were prepared using the phase inversion composition method (Solè et al., 2006; Wang et al., 2008). In order to guarantee the occurrence of minimum tension during the mixture of the components, the phase diagram of a system composed of dodecane, NaCl, iso-pentanol, SDS and water was previously built (Rahn-Chique et al., 2012a; Rahn-Chique et al., 2012b). For this study, 0.95 grams of an 11.5% wt/wt aqueous solution of NaCl were mixed with 0.846 gr of oil, and 0.2 gr of SDS. Then, enough iso-pentanol was added to reach 6.5% wt/wt in the total mixture. Following, the mixture was stirred with a mechanical stirrer at a velocity of 14.500 r.p.m. while additional water was added at a rate of 8 – 10 cm$^3$/s. During this lapse of time (1 minute), the agitator was moved conveniently to guarantee the correct homogenization of the mixture. Following this procedure, O/W nanoemulsions with average diameters between 401 and 440 nm were obtained.

The composition of the oil was systematically varied in order to produce systems with small ripening rate and/or neutral buoyancy. To these aims, previous experiments with non-ionic surfactants (Cruz, 2012) and hexadecane (García-Valero, 2011) showed that mixtures of dodecane with at least 7% wt/wt of squalene (SQ), or 30% wt/wt of tetrachloroethylene (TCE), respectively, were required. Since the addition of TCE promotes a considerable degree of ripening (see below), and the addition of squalene lowers the density of the mixture considerably, it was necessary to make a compromise in order to reach a system of maximum stability: $C_{12}$ 56 % wt/wt, TCE 23% wt/wt, SQ 21% wt/wt (Nano D). The nominal compositions of the mother emulsions used and their physical properties are given in Tables 1 and 2.

Due to the detection limit of the instruments, two different dilutions were necessary in order to follow the evolution of the systems (Table 3). In all dilutions, the final concentrations of NaCl and SDS were adjusted to 5 mM and 8 mM, respectively. The evolution of the average radius as a function of time was measured using a Brookhaven Goniometer at $n_0 = 4 \times 10^9$ part/cc. Electrophoretic measurements (Delsa 440SX, Beckman-Coulter), and turbidity evaluations as a function of height (Quickscan, Formulaction) were carried out at $n_0 = 4 \times 10^{10}$ part/cc. The latter measurements were undertaken at 200 acquisitions per hour during eight days. The creaming rate was calculated with the Migration software (Formulaction, 2002) employing Eq. (18):

$$V_g = \frac{2 R_i^2 \Delta\rho g}{9 \eta} \left[ |1-\phi| \Big/ \left|1 + 4.6\phi/(1-\phi)^3\right| \right] \tag{18}$$

*II. 3. Evaluation of the flocculation rate for doublet formation ($k_{11}$)*

To select a convenient wavelength of light for turbidity measurements, the absorption spectra of each emulsion, and each emulsion component was measured separately between 200 and 1100 nm using a UV-Visible spectrophotometer (Turner SP 890). The optimum wavelength for scattering studies is the one that guarantees negligible adsorption of all components, and shows a significant variation of the absorbance as a function of time. For these studies a value of $\lambda = 800$ nm was selected.

The appropriate particle concentration for aggregation studies was established from plots of $(dAbs/dt)_0$ vs. $n_0$. For this purpose the mother emulsions were diluted to produce initial particle concentrations ($n_0$) between $8 \times 10^9$ and $6 \times 10^{10}$ drop/cc (in the case of system A, the higher limit was extended to $2 \times 10^{11}$ drop/cc). The absorbance of each emulsion was measured at least thrice during ($t_{max} =$) 60 s after the addition of 600 mM of [NaCl], following the procedure previously described by Rahn-Chique (2012a, 2012b). From these curves, the initial slope of the absorbance, $(dAbs/dt)_0$, was computed. An approximately linear region of $(dAbs/dt)_0$ vs. $n_0$ was found below $6 \times 10^{10}$ part/cc.

Using a fixed particle concentration of $n_0 = 4.0 \times 10^{10}$ part/cc, the absorbance of each emulsion (A, B, C and D) was evaluated as a function of time, at different electrolyte concentrations (350 – 600 mM [NaCl]). One mother emulsion of each type (A, B, C, D) was synthesized for each salt concentration. The average radius of the drops and the polidispersity of each type of emulsion were quite reproducible: A (407 – 430 nm; 19 – 22 %), B (407 – 414 nm; 20 – 22%), C (401 – 443 nm; 19 – 21%), and D (400 – 450 nm; 25 – 27%). The appropriate amount of salt was introduced directly into the sample cell of the spectrophotometer (Turner, SP 890) by injecting 0.6 cc of brine to 2.4 cc of each diluted emulsion.

## III. RESULTS AND DISCUSSION

According to LSW the rate of Ostwald ripening is given by a linear slope of $R_a^3$ vs. t ($V_{exp}$ in Eq.(17)). However our simulations (Urbina-Villalba et al., 2009; 2012a; 2012b) suggest that flocculation and coalescence contribute significantly to $V_{exp}$. Thus $V_{exp} = V_{FCOB}$ (where F, C, O and B stand for Flocculation, Coalescence, Ostwald ripening, and Buoyancy). If a repulsive potential exists between the drops, the calculations predict a concave downward curve whose slope decreases with time approaching the LSW limit. The curve oscillates above and below its average slope because it results from two opposing trends: 1) the increase of the average radius of the emulsion produced by the elimination of drops either by coalescence and/or complete dissolution; and 2) the decrease of $R_a$ by the *molecular exchange* between the drops, which only *lowers* the average radius. Consequently, the curve of $R^3$ vs. t shows a saw-tooth variation in which $R_a$ grows when the number of drops diminishes, but it decreases at a constant number of drops.

Due to their small size, nanoemulsion drops are expected to be non-deformable. In the present case the drops are also highly charged owed to the adsorption of SDS. The electrostatic surface potential of the drops *increases* in the order C < A < B < D (Table 2) evidencing that the composition of the oil affects the adsorption process. Consequently, their flocculation rates are expected to *decrease* in the following order: C > A > B > D. Besides, since aggregation is necessary step prior to coalescence, both flocculation and coalescence are restricted for these systems in the absence of salt. Incidentally the highest value of $V_{exp}$ corresponds to system C, and its lowest variation to system D (see Figure 1).

According to Eq. (2) the ripening rate of dodecane in the stationary regime should be equal to 1 x $10^{-26}$ m$^3$/s. Figure 1 illustrates the long-time behavior of the systems. It is observed that the value of R$^3$ oscillates as a function of time. If the average slope of each curve is subtracted from R$^3$(t), a histogram of the deviations can be calculated. The frequency of the deviations can then be adjusted to a Gaussian distribution. The standard deviation of the distribution quantifies the amplitude of the oscillations which is found to decrease in the following order: A > C > D $\geq$ B (Figure 2). Thus, the addition of SQ to systems A and C (systems B and D, respectively), decreases the amplitude of the oscillations. Moreover, the average slope of systems A and C also decrease (Table 3). These results can be justified in terms of the lower solubility of the oil mixture which decreases with the addition of SQ. In fact, increasing the solubility of C$_{12}$ (system A) by the addition of TCE (system C) should increase V$_{exp}$, as it was experimentally found (Table 3). This does not occur when we move from system B to system D, because in this case the solubility increases with the addition of TCE, but also decreases with the augment of SQ (from 7 % to 23% (Table 1)).

So: Is the variation of $R_a^3$ vs. t a consequence of the degree of flocculation of the emulsions, or is it the result of the Ostwald ripening phenomenon? The truth is that when several destabilization processes are combined, V$_{exp}$ is neither a sound measure of flocculation (see below) nor a good measure of the ripening rate. In fact, system D shows a *negative* slope (-3.1 x $10^{-27}$ m$^3$/s) that cannot be explained by LSW theory, but is perfectly consistent with our previous simulations: if the rate of elimination of the drops is substantially decreased by lowering both the coalescence rate and the dissolution of the particles, only molecular exchange survives. In this case the simulations predict a decrease of the average radius as a function of time that LSW is unable to justify. In our view, this information along with the fact, that the zeta potential of systems A and B is very similar (Table 2) but their behavior is distinct, indicates that the Ostwald ripening phenomenon predominates: $V_{exp} \approx V_{OB}$.

In regard to buoyancy, Figure 2 confirms that despite the small size of their drops, the systems with the lower densities (A and B) are subject to a substantial degree of creaming. The base of the container clarifies as a function of time, showing higher transmittances, and the top is progressively obscured. Instead, the systems with TCE (systems C and D) do not show a significant amount of creaming. Notice the very different behaviors of these two systems in regard to $V_{exp}$ (Figure 1), despite their similarities with respect to $V_g$ (Figure 3 and Table 3). Figures 1 and 3 demonstrate that $V_{exp}$ and $V_g$ show trends that *contradict each other* in regard to emulsion stability. In particular, the creaming rate *is not* a convenient measurement of emulsion stability because *creaming is not the main destabilization phenomenon in the present systems*. In our understanding, the most unstable system is the one which presents the quickest departure from its initial state (and vice-versa). This is not necessarily equivalent to the fastest variation of the macroscopic properties. Moreover, such change can be caused by a predominant destabilization phenomenon, or a combination of several contributing phenomena. The system with the highest $V_{exp}$ rate (C) will only show the highest creaming rate if $\Delta\rho$ is appreciable, and this is not the case.

Figure 4 shows the variation of the initial slope of the absorbance as a function of the ionic strength. This slope is proportional to the initial aggregation rate: $k_{11}$ (in Eq. (12)), but also depends on the optical cross sections of singles and doublets (at constant $n_0$). In what respect to $(dAbs/dt)_0$ each set of data can be divided into two linear regions which in the case of solid particles correspond to diffusion limited cluster aggregation (DLCA) and reaction limited cluster aggregation (RLCA). The inflexion point is located around 400 mM [NaCl] which identifies the critical coagulation concentration (CCC). Above this salt concentration, the surface charge of the drops is screened by the electrolyte, and a maximum aggregation rate is achieved. Below 400 mM [NaCl] electrostatic interactions dominate.

As it is evident from Figure 5, $\sigma_1$ and $\sigma_2$ are sensitive functions of the average radius of the emulsions (Table 2). Within the DLCA regime, systems A and C show maxima for minimum values of $R_a$ and vice-versa. Since the average radius of the emulsions was measured in the absence of salt, it appears to be coincidental that the emulsions of these two systems show similar dependences of $k_{11}$ with respect to the salt concentration. Instead, a smooth variation of $k_{11}$ vs. [NaCl] is exhibited by the systems containing squalene (B and D) in the same range of salt concentrations. It is also remarkable that the curves of $k_{11}$ vs. [NaCl] are not straight lines within the RLCA regime. All systems show a change of slope around 370 - 385 mM (see also Fig. 6 and related discussion in Ref. [Rahn-Chique et al., 2012a]).

Theoretically the most unstable system with regard to aggregation is the one with highest $k_{11}$, lowest CCC, and highest $dk_{11}/d[NaCl]$ (for [NaCl] < CCC). All systems studied show a CCC of 400 mM. However, for the DLCA regime the absolute values of $k_{11}$ decrease in the order: C >> D $\approx$ B $\geq$ A, while for the RLCA regime they vary as: B > A > D > C. In regard to the change of $k_{11}$ with respect to the ionic strength ($dk_{11}/d[NaCl]$): B < A < D < C. In our view, the absolute value of $k_{11}$ is a determining factor in regard to aggregation, and therefore, the relative stability of these systems depends on the salt concentration.

In any event, since the change of $R_a^3$ vs. t is the result of all destabilization phenomena, it appears to be the most reliable measure of stability in these nanoemulsions.

V. ACKNOWLEGEMENT

The authors acknowledge the funding of the Agencia Española de Cooperación Internacional para el Desarrollo (AECID) through grant No. A/024004/09.

VI. BIBLIOGRAPHY

Cruz, E. (2012) Influencia de la Flotabilidad de las Gotas y el Fenómeno de Maduración de Ostwald sobre la Estabilidad de las Emulsiones Dodecano/Agua. MSc. Thesis, IVIC.

Evans F, Wennerstrom H, The Colloidal Domain: Where Physics, Chemistry, Biology and Technology Meet, VCH Publishers, 1$^{st}$ Ed., New York, 1994.

Finsy R (2004) On the critical radius of Ostwald ripening. *Langmuir 20*: 2975-2976.

Formulaction SA (2002) 10 Impasse Borde Basse 31240 L'Union – France. Website: www.formulaction.com

García-Valero N (2012), Estudio Comparativo de la Dinámica de Agregación de Nanoemulsiones Aceite/Agua y Suspensiones de Nanopartículas. Doctoral Thesis. IVIC.

Kerker M (1969) The scattering of light, and other electromagnetic radiation. Academic Press. New York. USA.

Kamogawa K, Matsumoto M, Kobayashi T, Sakai T, Sakai H, Abe M (1999) Dispersion and stabilizing effects of n-Hexadecane on Tetralin and Benzene metastable droplets in surfactant-free conditions. *Langmuir 15*, 1913-1917.

Lifshitz I.M, Slezov V.V. (1961) The Kinetics of Precipitation from Supersaturated Solid Solutions, *J. Phys. Chem. Solids 19*, 35-50.

Lips A, Smart CE, Willis E (1971) Light scattering studies on a coagulating polystyrene latex. *J. Chem. Soc. Faraday Trans. I 67:* 2979–2988.

Lips A, Willis E (1973) Low angle light scattering technique for the study of coagulation. *J. Chem. Soc. Faraday Trans. 69:* 1226–1236.

Maroto JA, de las Nieves FJ (1997) Estimation of kinetic rate constants by turbidity and nephelometry techniques in a homocoagulation process with different model colloids. *Colloid Polym. Sci. 275:* 1148–1155.

Mendoza, Y (2012), Bachelor in Science Thesis, IVIC.

Puertas A, Maroto JM, de las Nieves FJ (1998) Theoretical description of the absorbance versus time curve in a homocoagulation process. *Colloids Surf. A: Physicochem. Eng. Asp. 140:* 23–31.

Puertas AM, de las Nieves FJ (1997) A new method for calculating kinetic constants within the Rayleigh-Gans-Debye approximation from turbidity measurements. *J. Phys.: Condens. Matter 9:* 3313–3320.

Rahn-Chique K, Puertas AM, Romero-Cano MS, Rojas C, Urbina-Villalba G (2012a) Nanoemulsion stability: Experimental evaluation of the flocculation rate from turbidity measurements. *Adv. Colloid Interface Sci. 178*, 1-20.

Rahn-Chique K, Puertas AM, Romero-Cano MS, Rojas C, Urbina-Villalba G (2012b) Evaluación de la velocidad de floculación de nanoemulsiones aceite/agua. 2. Predicción de la turbidez de una dispersión dodecano/agua estabilizada con dodecil sulfato de sodio. *Interciencia 37*, 582-587.

Solè I, Maestro A, González C, Solans C, Gutiérrez JM (2006) Optimization of nanoemulsion preparation by low-energy methods in an ionic surfactant system. *Langmuir 22:* 8326–8332.

Urbina-Villalba G, Forgiarini A, Rahn K, Lozsán A (2009) Influence of flocculation and coalescence on the evolution of the average radius of an o/w emulsion. Is a linear slope of r3 vs. t an unmistakable signature of ostwald ripening? *Phys. Chem. Chem. Phys. 11:* 11184–11195.

Urbina-Villalba G, García-Sucre M (2000) Brownian dynamics simulation of emulsion stability. *Langmuir 16:* 7975–7985.

Urbina-Villalba, G (2012a) El Fenómeno de Maduración de Ostwald. Predicciones de las Simulaciones de Estabilidad de Emulsiones sobre la Evolución del Radio Cúbico Promedio de una Dispersión Aceite/Agua. ArXiv: 1303.2097 (http://arxiv.org/abs/1303.2097)

Urbina-Villalba, G, Rahn-Chique, K (2012b) Short-Time evolution of alkane-in-water nano-emulsions. ArXiv:1303.1423 (http://arxiv.org/abs/1303.1423)

von Smoluchowski M (1917) Versuch einer mathematischen theori der koagulationskinetik kolloider losungen. *Z. Phys. Chem. 92:* 129–168.

Wang L, Mutch KJ, Eastoe J, Heenan RK, Dong J (2008) Nanoemulsions prepared by a two-step low-energy process. *Langmuir 24:* 6092–6099.

Wagner C. (1961) Theorie der Alterung von Niederschlägen durch Umlösen, *Z. Elektrochemie 65*, 581-591.

Table 1: Nominal Composition of the Emulsions

| System / Composition | A | B | C | D |
|---|---|---|---|---|
| SDS (%wt) | 1.5 | 1.5 | 1.5 | 1.5 |
| NaCl (%wt) | 0.94 | 0.94 | 0.94 | 0.94 |
| Isopentanol (%wt) | 1.04 | 1.04 | 1.04 | 1.04 |
| Weight fraction of water | 0.93 | 0.93 | 0.93 | 0.93 |
| Weight fraction of oil | 0.07 | 0.07 | 0.07 | 0.07 |
| Weight Percentage of Dodecane | 100 | 93 | 70 | 56 |
| Weight Percentage of Squalene | --- | 7 | --- | 23 |
| Weight Percentage of Tetrachloro-ethylene | --- | --- | 30 | 21 |

Table 2: Properties of the mother emulsions

| System / Composition | A | B | C | D |
|---|---|---|---|---|
| Density of the oil mixture (g/ml) | 0.750 ± 0.001 | 0.756 ± 0.001* | 0.991 ± 0.001* | 0.955 ± 0.001* |
| Refractive Index of the oils | 1.425 ± 0.002 | 1.430 ± 0.002 | 1.449 ± 0.002 | 1.450 ± 0.002 |
| Average Radius of the drops (nm) | 400 ± 100 | 420 ± 90 | 410 ± 70 | 400 ± 100 |
| $N_0$ (particles/ml) | $2.2 \times 10^{+12}$ | $2.2 \times 10^{+12}$ | $1.7 \times 10^{+12}$ | $8.3 \times 10^{+12}$ |
| Zeta Potential of the Drops (mV) | -84 ± 14 | -89 ± 14 | -73 ± 17 | -94 ± 14 |
| $C_{RGD}$ | 0.313 | 0.330 | 0.320 | 0.396 |
| $\alpha$ | 2.19 | 2.19 | 2.19 | 2.19 |

*These errors were formerly estimated by weighting five 1-ml samples of the mixtures. However, such a procedure only validates the precision of the measurements. Since the provider (Aldrich) can only assess the exactness of the density of pure dodecane up to three significant figures, this estimation of the error (0.001) was employed in all cases.

Table 3: Composition and Properties of diluted Nanoemulsions

| System / Composition | A | B | C | D |
|---|---|---|---|---|
| SDS (M) | $8 \times 10^{-3}$ | $8 \times 10^{-3}$ | $8 \times 10^{-3}$ | $8 \times 10^{-3}$ |
| NaCl (M) | $5 \times 10^{-3}$ | $5 \times 10^{-3}$ | $5 \times 10^{-3}$ | $5 \times 10^{-3}$ |
| $N_0$ (particles/ml) for Turbidity measurements | $4 \times 10^{+9}$ | $4 \times 10^{+9}$ | $4 \times 10^{+9}$ | $4 \times 10^{+9}$ |
| $N_0$ (particles/ml) for the evaluation of the Creaming velocity | $4 \times 10^{+10}$ | $4 \times 10^{+10}$ | $4 \times 10^{+10}$ | $4 \times 10^{+10}$ |
| Creaming Velocity (cm/s) | $(3.6 \pm 0.5) \times 10^{-6}$ | $(2.1 \pm 0.5) \times 10^{-6}$ | $\sim 0$ | $(-3.4 \pm 0.5) \times 10^{-7}$ |
| $d\overline{R}^3/dt$ ($m^3/s$) | $(2.7 \pm 0.9) \times 10^{-26}$ | $(1.6 \pm 0.9) \times 10^{-28}$ | $(4.7 \pm 0.2) \times 10^{-26}$ | $(-3.1 \pm 0.8) \times 10^{-27}$ |

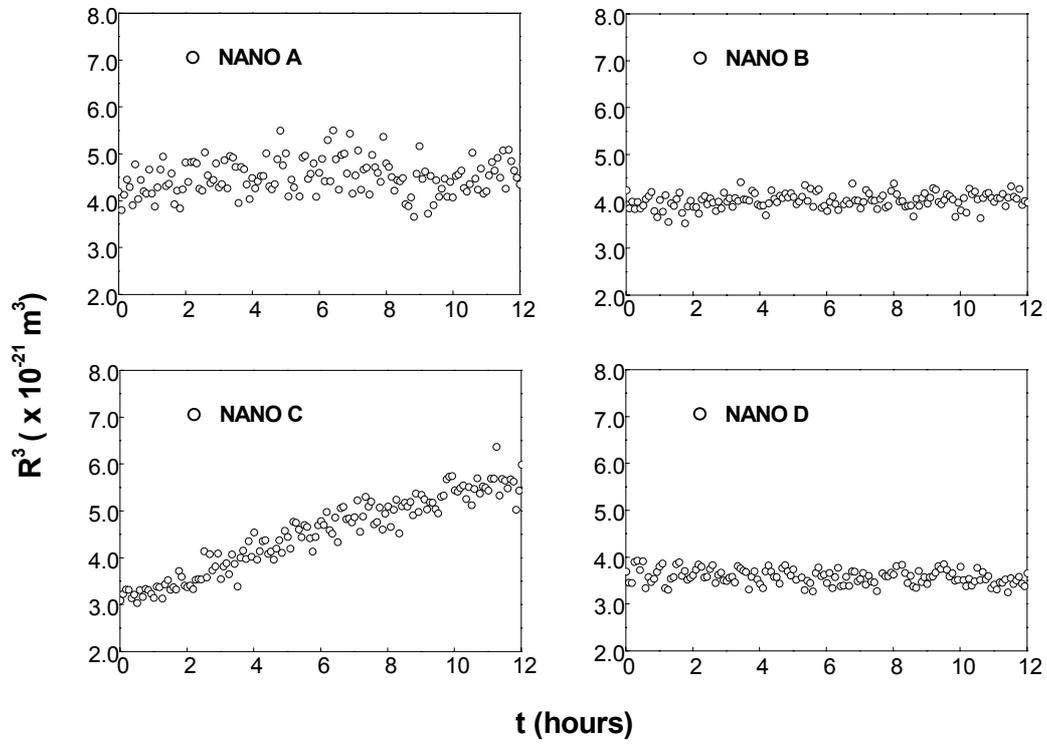

FIGURE 1: Change of $R_a^3$ vs. t for each nanoemulsion.

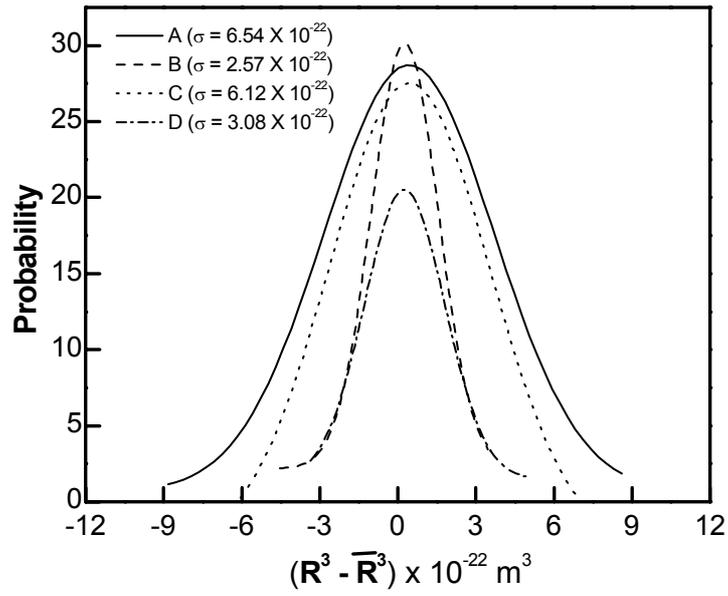

FIGURE 2: Frequency of the deviations of $R_a^3$ vs. t with respect to its average slope.

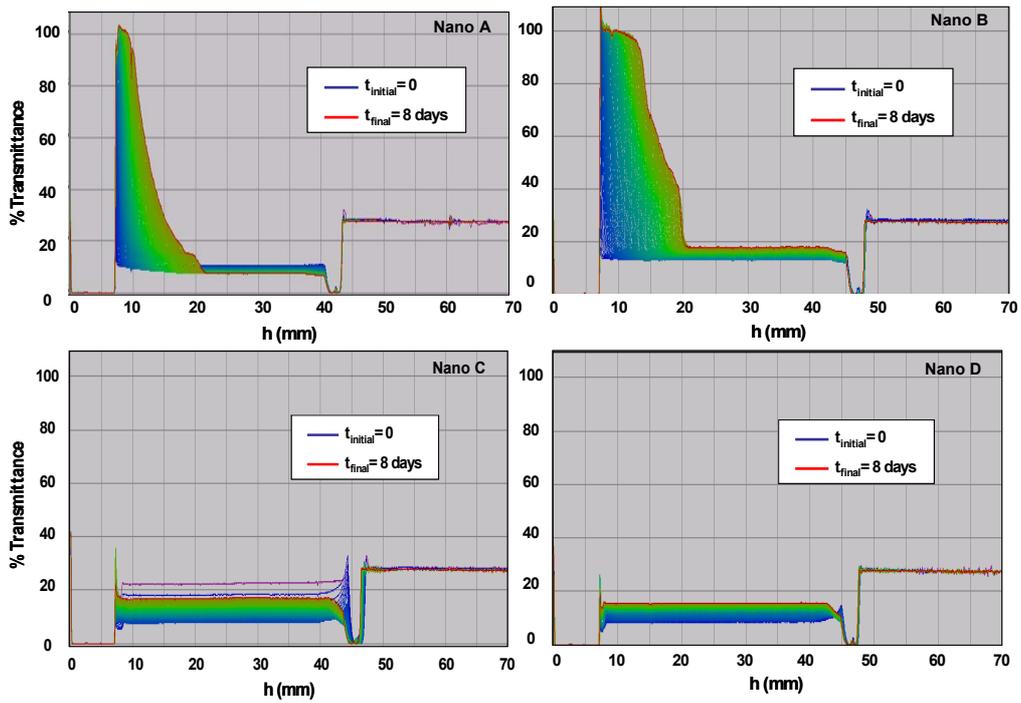

FIGURE 3: Change in Transmittance of the mother emulsions as a function height and time.

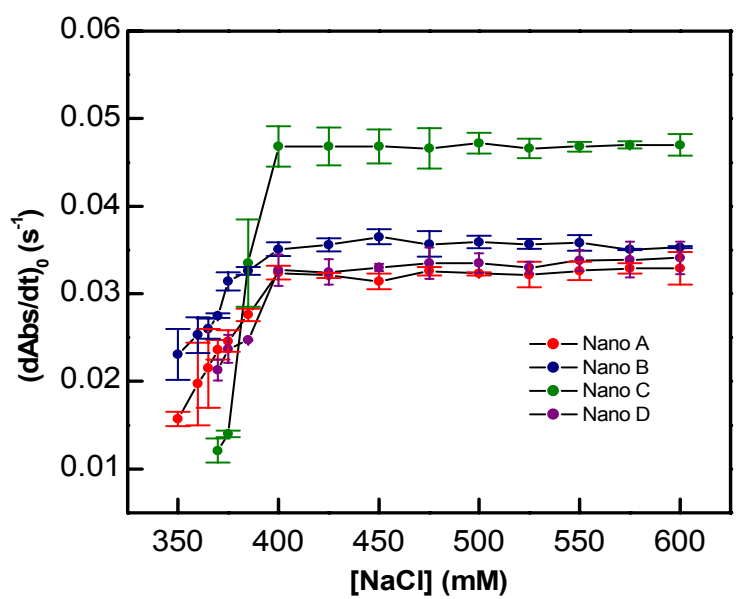

FIGURE 4: Initial slope of the absorbance as a function of the ionic strength for systems A, B, C and D.

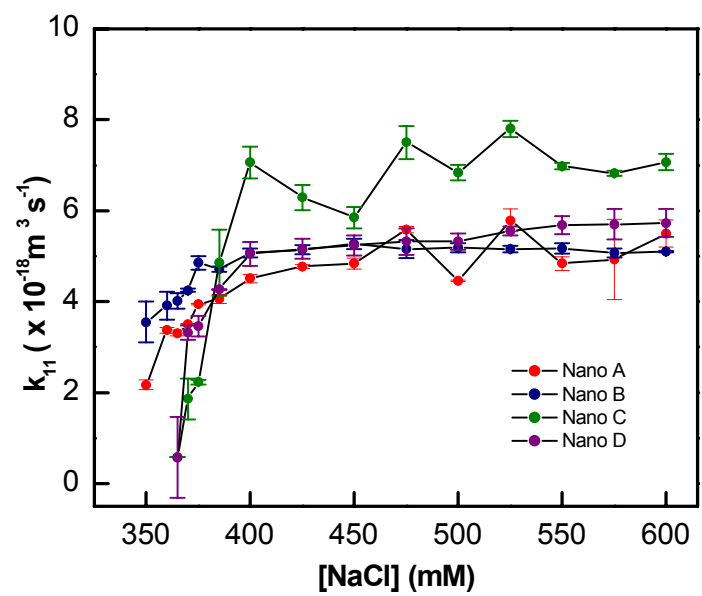

FIGURE 5: Dependence of $k_{11}$ with the salt concentration for systems A, B, C and D.

TABLE CAPTIONS

TABLE 1: Nominal Composition of the Emulsions

TABLE 2: Properties of the mother emulsions

TABLE 3: Composition and Properties of diluted Nano emulsion

FIGURE CAPTIONS

FIGURE 1: Change of $R_a^3$ vs. t for each nanoemulsion.

FIGURE 2: Frequency of the deviations of $R_a^3$ vs. t with respect to its average slope.

FIGURE 3: Change in Transmittance of the mother emulsions as a function height and time.

FIGURE 4: Initial slope of the absorbance as a function of the ionic strength for systems A, B, C and D.

FIGURE 5: Dependence of $k_{11}$ with the salt concentration for systems A, B, C and D.